# NEW RESULTS FOR REACTION RATE OF THE PROTON RADIATIVE CAPTURE ON $^3$H

S. B. DUBOVICHENKO[1,2,*], A. V. DZHAZAIROV-KAKHRAMANOV[1,2,†]
and N. V. AFANASYEVA[2,‡]

[1]*Fesenkov Astrophysical Institute "NCSRT" ASC MID Republic of Kazakhstan (RK), 050020, Observatory 23, Kamenskoe plato, Almaty, RK*
[2]*Institute of nuclear physics ME RK, 050032, str. Ibragimova 1, Almaty, RK*
[*]dubovichenko@mail.ru, [†]albert-j@yandex.ru
[‡]n.v.afanasyeva@gmail.com

## ABSTRACT

Calculations of the reaction rate of the proton radiative capture on $^3$H at temperatures from 0.01 $T_9$ up to 5 $T_9$, which are based on the theoretical results for the astrophysical $S$-factor and take into account the latest experimental data, were carried out. Theoretical results for the $S$-factor at energies from 1 keV up to 5 MeV were obtained in the framework of the modified potential cluster model with the classification of orbital states according to Young tableaux. On the basis of used nuclear model of the interaction of p and $^3$H particles there was shown possibility of description the latest experimental data for the $S$-factor at the energy range from 50 keV up to 5 MeV.

## 1. INTRODUCTION

The proton capture on $^3$H reaction is of interest from both theoretical and experimental points of view for understanding the dynamics of photonuclear processes involving the lightest atomic nuclei at low and ultralow, i.e., astrophysical energies [1]. It also plays a role in the nucleosynthesis of primordial elements in the early Universe [1,2,3] leading to the pre-stellar formation of $^4$He nuclei. Therefore, experimental studies of this reaction continue. New data for the total cross section of proton radiative capture on $^3$H and the astrophysical $S$-factor in the energy range from 50 keV to 5 MeV [4] and at 12 and 39 keV [5] in the center of mass system (c.m.) have been obtained. These data will be used by us for further comparison with the calculation results.

Upon cooling to a temperature of ~0.8 MeV, the processes of the primordial nucleosynthesis became possible [6,7] with the formation of stable $^2$H, $^3$He and $^4$He nuclei and, also stable in the first minutes of the Universe, the $^3$H nucleus. These reactions are shown in Table 1 – the processes of the radiative capture are marked by italic. In table also the data of the $S$-factors and total cross sections at low energies in the energy range 10 – 20 keV were given with references to original works with these results. Table 1 shows that only one of these reactions, No.4, results in energy absorption 0<Q. All of the others lead to energy release Q>0. Some inverse nuclear reactions, for example, photodisintegration of $^{3,4}$He and $^{2,3}$H by gamma-quantum cannot occur because of their extremely low energies at which weak processes cannot keep the balance [7].

Therefore the constant synthesis of stable nuclei without their further disintegration to lighter nuclei becomes possible.

This was the situation when the Universe was about 100 sec old and the number of protons and neutrons was comparable – approximately 0.2 neutrons to each proton. The epoch of primordial nucleosynthesis finished at approximately 200 sec [6] by which time practically all neutrons are bound into $^4$He nuclei and the number of $^4$He is about 25% of the number of $^1$H nuclei. At that point the content of $^2$H and $^3$He relative to $^1$H was about $10^{-4}$–$10^{-6}$ [1–3,7].

Table 1. Basic reaction of the primordial nucleosynthesis with light nuclei.

| No. | Process | Released energy in MeV | Astrophysical $S$-factor in keV b at 10 – 20 keV in center of mass – the accurate energy is stated in square brackets | The total cross section $\sigma_t$ in µb for the given energy | Reference |
|---|---|---|---|---|---|
| 1. | $p+n \to {}^2H+\gamma$ | 2.225 | $3.18(25)\cdot 10^{-3}$ [10.0] | $3.18(25)\cdot 10^2$ [10.0] | [8] |
| 2. | $^2H+p \to {}^3He+\gamma$ | 5.494 | $3.0(6)\cdot 10^{-4}$ [10.4] | $1.0(2)\cdot 10^{-2}$ [10.4] | [9] |
| 3. | $^2H+n \to {}^3H+\gamma$ | 6.257 | $1.2\cdot 10^{-5}$ [10.5]* | 1.1 [10.5]* | [10] |
| 4. | $^3H+p \to {}^3He+n$ | –0.763 (see [11]) | 2536 [12]*** | 81537 [roughly at 12 keV above the threshold or 1.03354 MeV in l.s.] | [12] |
| 5. | $^3He+n \to {}^3H+p$ | 0.764 | 63.2 [10.3] | $6.14(16)\cdot 10^6$ [10.3] | [13] |
| 6. | $^3H+p \to {}^4He+\gamma$ | 19.814 | $2.2\cdot 10^{-3}$ [10.0] | $4.0\cdot 10^{-2}$ [10.0] | [5] |
| 7. | $^3He+n \to {}^4He+\gamma$ | 20.578 | $1.7\cdot 10^{-4}$ [18.4] | 9.2(2.0) [18.4] | [14] |
| 8. | $^2H+{}^2H \to {}^3He+n$ | 3.269 | 51.4(2.0) [9.94]<br>53.05(0.55) [10.0]*** | 241.3(9.4) [9.94]**<br>255.1(2.9) [10.0] | [15]<br>[16] |
| 9. | $^2H+{}^2H \to {}^3H+p$ | 4.033 | 56.1(1.6) [9.97] | 270.4(7.6) [9.97] | [17] |
| 10. | $^2H+{}^3He \to {}^4He+p$ | 18.353 | 7480(200) [10.7] | 0.5(1) [10.7]** | [18] |
| 11. | $^2H+{}^3H \to {}^4He+n$ | 17.589 | 12328.4 [9]*** | 14200 [9] | [19] |
| 12. | $^2H+{}^2H \to {}^4He+\gamma$ | 23.847 | $5.7(2.4)\cdot 10^{-6}$ [10.0] | $2.9(1.2)\cdot 10^{-5}$ [10.0] | [20] |
| 13. | $^2H+{}^3He \to {}^5Li+\gamma$ | 16.66 | 0.41 [111]*** | 5.3 [111] | [21] |
| 14. | $^2H+{}^3H \to {}^5He+\gamma$ | 16.792 | 0.17 [90]*** | 50 [90] | [21] |

\* - theoretical value calculated on the basis of the Modified Potential Cluster Model
\*\* - the value calculated on the basis of the $S$-factor
\*\*\* - the value calculated on the basis of the total cross section

Thus $^4$He was the last nucleus to emerge at the initial stage of nucleosynthesis because heavier nuclei such as C and O could only be synthesized in the process of nuclear reactions in stars. The reason for this is the existence of some an instability gap for light nuclei ($A = 5$), which, apparently, can't be bridged in the process of initial nucleosynthesis. In principle, $^4$He could have given rise to heavier nuclei ($A = 7$) in the $^4$He + $^3$H → $^7$Li + γ and $^4$He + $^3$He → $^7$Be + γ reactions. However the Coulomb barrier for these reactions is about 1 MeV while the kinetic energy of the nuclei at temperatures of ~ 1 $T_9$ is of the order of 0.1 MeV and probability of such reactions will be negligible [22]. The mechanism of synthesis of $^4$He explains its abundance in the Universe confirms its origin at the pre-stellar stage and corroborates the Big Bang theory.

It is important to estimate the *S*-factors of reactions 1–14. For example, as will be seen further, the astrophysical *S*-factor of proton capture on $^2$H at an energy of 1 keV is in 5–10 times lower than the *S*-factor of the proton capture on $^3$H at the same energy [10]. This means that the latter process, which contributes to the formation of $^4$He in primordial nucleosynthesis, is much more likely, in spite of the lower abundance of $^3$H relative to $^2$H [6,7,23]. Most data available in the literature [1–4,10,22] relate to the abundance of elements such as $^3$He at present time. This is generally confirmed by modern astrophysical observations [6,23]. However, the abundance of $^3$H for the first 100–200 s after the Big Bang cannot be much smaller than that of $^2$H since the neutron capture reaction, in spite of the reduction of neutron numbers down to 0.2 of the proton numbers, can go on deuteron at any energy. In addition, the half-life of $^3$H is 4500(8) days [24] and do not make a real contribution to the decrease of the number of $^3$H at the first few minutes after the Big Bang.

The quantity of tritium, except in process No.3, increases due to reactions No.5 and No.9, but can decrease due to processes No.6 and 11. At energies lower than 0.8 MeV reaction No.4 makes virtually no contribution to reductions in tritium. Meanwhile, the total cross section of reaction No.11 is about 14.2 mb at 9 keV [19] and of the reaction No.6 is about $4·10^{-2}$ μb at 10 keV [5] show their small relative contributions to the formation of $^4$He. However the number of deuterons available for reaction No.11 is approximately 4–5 orders of magnitude less than the number of protons taking part in reaction No.6. Therefore the overall contribution of the two reactions in pre-stellar formation of $^4$He will be similar.

Reaction No.12 proceeds with comparatively low probability, since the *E*1 process is forbidden by the isospin selection rules. This leads to the factor $\left(Z_1/m_1^J + (-1)^J Z_2/m_2^J\right)$ at multipolarity of γ-quantum of $J = 1$ [10]. This product defines the value of the total cross sections of the radiative capture and *E*1 processes with the same Z/m ratio, for particles of the initial channel leads to zero cross sections. The probability of the allowed *E*2 transitions in such processes is usually nearly 1.5 to 2.0 orders of magnitude less [25] that was noted earlier in reviews [6,7]. This fact well demonstrates cross section's data for reaction No.12 from Table 1, the value of which is lowest. For two last reactions No.13, 14 we cannot find experimental data at lower energies.

Let us show furthermore the reaction rates given in work [26] in the form of parametrizations. Shape of these rates for first reactions, leading to the formation of $^4$He or nuclei with mass of 3, is shown in Fig. 1. One can see that considered reaction is at the third level for rate of forming $^4$He and its rate in some times lower, for example, that the reaction rates of $^3$H(d,n)$^4$He or $^3$He(d,p)$^4$He. However, at the energy about 10 $T_6$ the rate of the last reaction equalized with the reaction rate of the proton radiative capture on $^3$H.

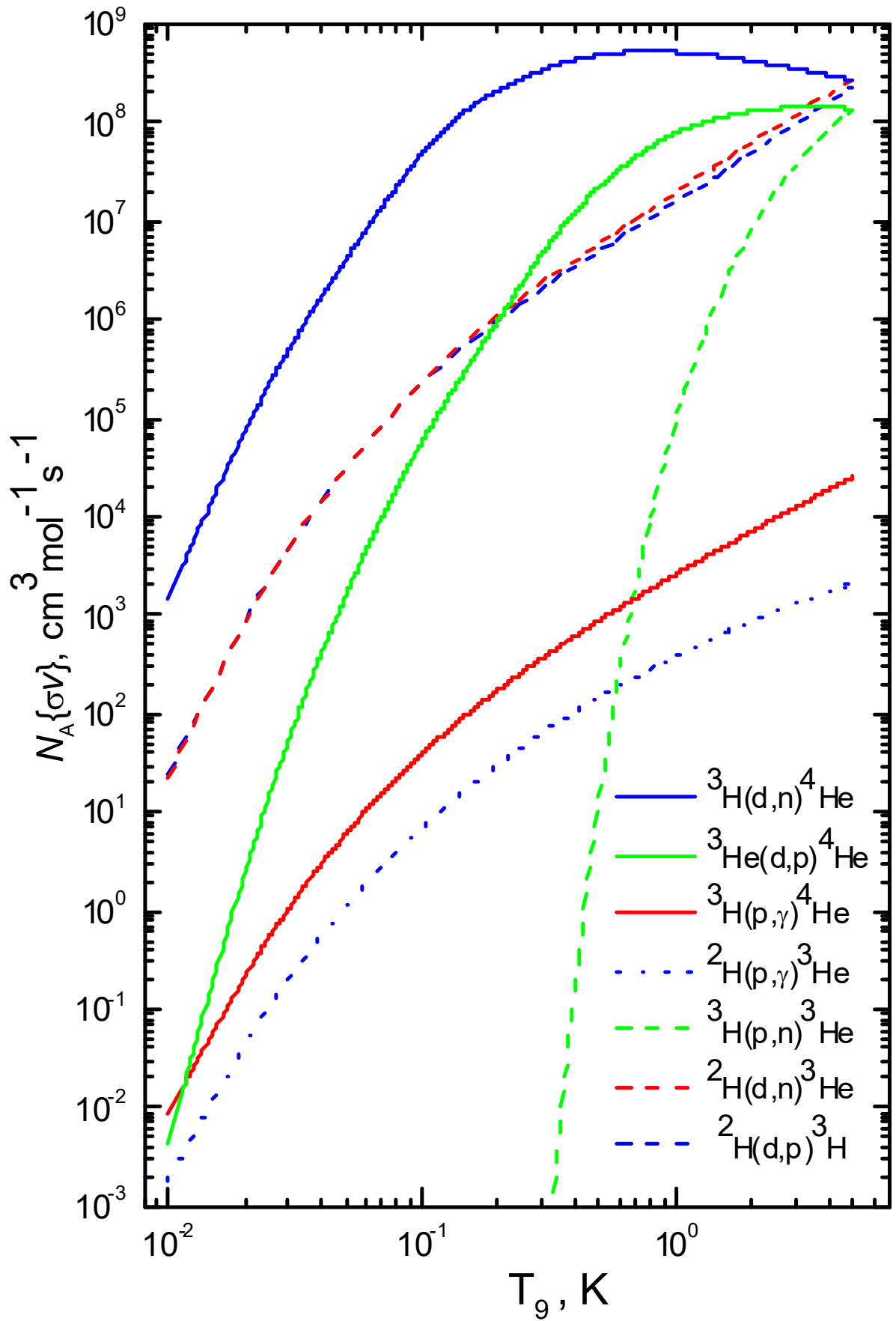

FIG. 1. Reaction rate from [26].

All these results and more new data from works [27,28] show that the contribution of the p($^3$H, γ)$^4$He capture reaction into the processes of primordial nucleosynthesis is relatively small. However, it makes sense to consider this process for making the picture complete of the formation of prestellar $^4$He and clearing of mechanisms of this reaction. In addition, as it was shown furthermore, our calculations of this reaction rate, based on the modern data of the astrophysical *S*-factors [5], lays slightly lower from the results of works [26,29–31]. The latest works do not take into account new data [4,5], which were taken into account by us in this work, and our results can be considered as an improved data on the rate of the considered reaction.

Moreover it should be noted that our understanding of the different stages in the formation of the Universe, of the processes of nucleosynthesis occurring in it and of the properties of new stars, is still developing. Therefore there is a pressing need to acquire new information on primordial nucleosynthesis and on the mechanisms of the Universe's formation and this is one of the main tasks for the construction of a unified cosmological model. All of this directly applies to the detailed study of the p$^3$H capture reaction in the astrophysical energy region on the basis of the modern nuclear model. This model, as shown below, has already demonstrated its efficiency in the description of the characteristics of almost 30 such reactions [10,32–37].

## 2. MODEL USED

In our earlier works [10,32–37] the possibility of describing astrophysical *S*-factors or total cross-sections of radioactive capture for three dozen processes on the basis of two-body potential cluster model (PCM) was demonstrated. Preliminary results [38,39] were obtained for p$^3$H-capture at astrophysical energies. The calculations of these reactions were carried out on the basis of the modified variant of PCM (MPCM) with forbidden states (FSs) and classification of states according to Young tableaux [40].

The well-defined success of the MPCM in the description of total cross sections of this type can be explained by the fact that the potentials of the intercluster interaction in the continuous spectrum are constructed on the basis of the known elastic scattering phase shifts or structure of the resonance spectrum levels of the final nucleus. For the discrete spectrum it is on the basis of the main characteristics of the bound states (BSs) of such nuclei: the excited (ES) or the ground (GS) states. These intercluster potentials are based also on the classification of the cluster states according to Young tableaux [41]. This enables the determination of the presence and quantity of the FSs in each partial wave. This means finding the number of wave function (WF) nodes in such cluster systems [32].

Such potentials enable us to carry out the calculations of some basic characteristics of particle interactions in the elastic scattering processes and reactions. These can be the astrophysical *S*-factors of radiative capture reactions [42] or the total cross sections of these reactions [43] and include radiative capture cross sections in the astrophysical and thermal energy range as discussed in [10,32–37]. Based on this approach we obtained accurate descriptions of the total cross sections of the radiative capture processes of almost thirty reactions for light nuclei at thermal and astrophysical energies.

Therefore, continuing the study of thermonuclear reactions on the basis of the MPCM [10,32] with separation of orbital states according to Young tableaux, let us consider the astrophysical *S*-factor of the radiative proton capture on $^3$H at energies of 1 keV–5 MeV and the rate of this reaction from $T_9 = 0.01$ to $T_9 = 5$. Preliminary results for the *S*-factor of this reaction at astrophysical energies in the frame of the MPCM were given in [38,39]. In the current work new results for the rate of the proton capture on $^3$H have been obtained and compared those in [38,39]. The potentials of the scattering processes and bound p$^3$H states are improved and detailed classification carried out for the orbital states of the p$^3$H system according to Young tableaux and isospin.

## 3. POTENTIALS FOR p$^3$H AND p$^3$He SYSTEMS

For calculations of the photonuclear processes in the system considered the nuclear part of the intercluster potential of p$^3$He interactions for each partial wave can be expressed in the form

$$V_{JLS\{f\}}(R) = V_0(JLS\{f\})\exp[-\alpha(JLS\{f\})R^2] + V_1(JLS\{f\})\exp[-\gamma(JLS\{f\})R] \quad (1)$$

with a point-like Coulomb term. This potential, as for that of the p$^2$H system [10,33], is constructed so as to correctly describe the corresponding partial phase shift of the p$^3$He elastic scattering [38,39].

Consequently the pure (with respect to isospin $T = 1$) potentials of p$^3$He interactions for the elastic scattering processes were obtained, and their parameters are listed in (2) and (3) [32,40]. The singlet potentials of the form given in expression (1) for the p$^3$He scattering, pure with respect to isospin $T = 1$ [32,40]:

$$p^3He\ System$$
$$^1S\ \text{wave} - V_0 = -110.0\ \text{MeV},\ \alpha = 0.37\ \text{fm}^{-2},\ V_1 = +45.0\ \text{MeV},\ \gamma = 0.67\ \text{fm}^{-1}, \quad (2)$$
$$^1P\ \text{wave} - V_0 = -15.0\ \text{MeV},\ \alpha = 0.1\ \text{fm}^{-2}. \quad (3)$$

Note that this singlet and pure (with respect to isospin $S$) phase shift of the p$^3$He elastic scattering is used further for calculation of the singlet p$^3$H phase shifts with the isospin $T = 0$. The singlet $^1P_1$ phase shift of the p$^3$He elastic scattering with $T = 1$ used in our calculations of the $E1$ transition to the ground state of $^4$He in the p$^3$H channel with $T = 0$. The scattering phase shifts obtained with such potentials are given in our previous papers [38,39].

The singlet, isospin and Young tableaux mixed $S$ phase shift of the elastic p$^3$H scattering, determined from the experimental differential cross sections, and used later for obtaining the pure p$^3$H phase shifts for potential (1) at $V_1 = 0$ with parameters

$$V_0 = -50\ \text{MeV},\ \alpha = 0.2\ \text{fm}^{-2}. \quad (4)$$

Then, the following parameters at $V_1 = 0$ for the pure p$^3$H potential with $T = 0$ in the $^1S$ wave in [38,39] have been found:

$$V_0 = -63.1\ \text{MeV},\ \alpha = 0.17\ \text{fm}^{-2}. \quad (5)$$

Scattering phase shifts for potentials (4) and (5) are given in papers [38,39]. The pure (according to Young tableaux) interactions thus obtained can be used for the calculation of different characteristics of the bound ground state $^4$He in the p$^3$H channel. The degree of agreement of the results obtained in this case with experiment now depends only on the degree of clusterization of this nucleus in the channel considered and here one supposes that this degree is high enough that the spectroscopic factor of the channel will be close to unity.

The interaction potential (5) obtained in [38,39] on the whole correctly describes the channel binding energy of the p$^3$H system (to several keV) and the root-mean-square radius of $^4$He. Using this potential and the potential of the $^1P$ scattering wave with the point-like coulomb term for the p$^3$H system from (3), the differential and total cross sections of proton radiative capture on $^3$H were calculated in [38,39] and [40] respectively. The astrophysical S-factors at energies down to 10 keV were also calculated.

It should be noted that at that time experimental data for the S-factor only was known in the energy region above 700–800 keV [44]. Subsequently new experimental data were obtained in [4] and [5]. It will be of interest to explore whether the potential cluster model with the singlet $^1P$ potential obtained earlier and the refined interaction of the pure ground $^1S$ state of $^4$He is capable of describing this new more accurate data.

Our preliminary results [38,40] have shown that for calculation of the S-factor at energies of the order of 1 keV the same conditions as in the p$^2$H system [10,33] should be satisfied. In particular the accuracy of values obtained for the binding energy of $^4$He in the p$^3$H channel should be increased. New modified programs as described in [32,33] were used in the current work in order to improve parameters of the potential of the ground state for the p$^3$H system of $^4$He as given in [43].

The results for pure potentials (with respect to isospin of $T = 0$) are given in the expression (1) with the following parameters:

$$p^3H \text{ System}$$
$$^1S \text{ wave} - V_0 = -62.906841138 \text{ MeV}, \quad \alpha = 0.17 \text{ fm}^{-2}. \quad (6)$$
$$^1P \text{ wave} - V_0 = +8.0 \text{ MeV}, \quad \alpha = 0.03 \text{ fm}^{-2}. \quad (7)$$

These results obtained in [39] differ from those presented in [38] by approximately 0.2 MeV. This difference is mainly connected with the use, in the new calculations, of more accurate values of masses of p and $^3$H particles [45] and more accurate description of the binding energy of $^4$He in the p$^3$H channel. Using this value of –19.813810 MeV was obtained [11]. The calculation with the potential considered here gives –19.81381000 MeV.

The behavior of the "tail" of the numerical wave function (WF) $\chi_L(R)$, for the p$^3$H system bound state at large distances, was verified using an asymptotic constant (AC) $C_W$ [46,47]

$$\chi_L(r) = \sqrt{2k_0} \, C_W W_{-\eta L+1/2}(2k_0 r) \quad (8)$$

where a value of $C_W = 4.52(1)$ was used for the reasons described in detail below; $\mu Z_1 Z_2 e^2/(q\hbar^2)$ is the coulomb parameter, where q is the wave number determined by the energy of interacting particles in the initial channel; L is the orbital moment; $W_{-\eta L+1/2}(2k_0 R)$ is the Whittaker function [47]; $k_0 = \sqrt{2\mu \frac{m_0}{\hbar^2} E}$ is the wave number of the GS; E is the binding energy in p$^3$H channel; and the constant $\hbar^2/m_0$ is equal to 41.4686 MeV fm$^2$ where $m_0$ is the atomic mass unit (amu) [32].

The reduced error of the asymptotic constant $C_W$ was achieved by averaging it over the range in which its variation is a minimum. To find the region for $C_W$ stabilization we calculated it starting at the maximum distances that were considered by us of about 20–30 fm. This stabilization usually occurs at distances of about 7–12 fm. In this region the $C_W$ changes least with a variation of about $10^{-3}$. At distances below the stabilization region for the WF $\chi_L(R)$ we use numerical values of this function obtained from the Schrödinger equation solution. At distances above the stabilization region it is calculated from its asymptotic (8) determined by the Whittaker function $W_{-\eta L+1/2}(2k_0r)$ using the value for $C_W$ found in the stabilization region.

The experimentally determined value of $C_W$ in [47] is 5.16(13) for the p$^3$H channel. For the n$^3$He system a value of 5.10(38) was obtained. This is very close to the constant of the p$^3$H channel for our GS potential from (6). [46] reports a value of 4.1 constant of the n$^3$He system and 4.0 for p$^3$H. The average value between the results of works [46] and [47] is in good agreement with our results for the GS potential from (6). Apparently, there is a considerable difference between the data of asymptotic constants. For the n$^3$He system reported values range from 4.1 to 5.5 and for the p$^3$H channel from 4.0 to 5.3.

The results reported in [48], where the average spectroscopic $S_f$ factor was 1.59 and the average value of the asymptotic normalizing coefficient $A_{NC}$ (ANC) was 6.02 fm$^{-1/2}$, were obtained on the basis of calculations with different potentials. The relationship between ANC and dimensional AC $C_W$ [43,49]:

$$A_{NC}^2 = S \times C^2 \qquad (9)$$

where $C$ can be found from

$$\chi_L(r) = CW_{-\eta L+1/2}(2k_0r) \qquad (10)$$

This dimensional constant is related to the non-dimensional $C_W$ used by us by $C = \sqrt{2k_0}C_W$. So using the values of ANC and $S_f$ given in [48] we obtained a value for $C$ of 4.77 fm$^{-1/2}$. In this case $\sqrt{2k_0} = 1.30$ so the dimensionless AC $C_W$ is 3.67. This is slightly less than values obtained here and given in [46,47]. However, if the spectroscopic factor determines the possibility of certain two-body channel, then it is unlikely to be more than unity. $S_f = 1.0$ would give a $C$ of 6.02 and $C_W$ of 4.63. This agrees acceptably with the given above dimensionless value of 4.52 for the $^1S$ potential of the ground state from (6).

For the charge radius of $^4$He, with the potential from (6), we have obtained a value of 1.78 fm (calculation methods for charge radius are described in [32,33,38,40]. The experimentally determined value of $^4$He radius 1.671(14) fm [11]. For these calculations we have used the values of the tritium radius of 1.73 fm from [24] and the proton radius of 0.8775 fm from data base [45].

## 4. ASTROPHYSICAL S-FACTOR AND REACTION RATE

For calculation of the astrophysical S-factor the usual expression was used (see, for example, [29])

$$S = \sigma E \exp(31.335 Z_1 Z_2 \sqrt{\frac{\mu}{E}}), \qquad (11)$$

where $\sigma$ is the total cross section in barn, $E$ is the center-of-mass energy in keV, $Z_i$ are the particle charges, $\mu$ is the reduced mass of particles in aum [1].

The total cross sections and the astrophysical *S*-factor of the proton radiative capture process on $^3$H have previously been calculated based on the modified potential cluster model [38]. It was assumed that the main contributions to the cross sections of *E*1 photodisintegration of $^4$He in the p$^3$H channel, or to the proton radiative capture on $^3$H, were isospin-flip transitions for which $\Delta T = 1$ [50]. Therefore our calculations will assume that the $^1P_1$ potential for p$^3$He scattering is pure with respect to the isospin ($T = 1$) singlet state of this system and that the $^1S$ potential for the ground state is pure with respect to the isospin $T = 0$ bound state of $^4$He in the p$^3$H channel [38].

Using these assumptions, the *E*1 transition with refined potential of the ground state of $^4$He, as shown in (6), was re-calculated. The results for the astrophysical *S*-factor at energies from 1 keV up to 5 MeV are shown in Fig. 2 by the green solid line. In particular the new results obtained for the energy region from 10 keV to 5 MeV are in close agreement with our previous results as given in [39].

Fig. 2 also shows the resulting values from new experimental data [4,5] and additional data from [51] not known to us earlier. It can be seen from this figure that the calculations performed about 20 years ago [38] well reproduce the data on the *S*-factor obtained in [4] at energies of p$^3$H capture from 50 keV to 5 MeV (c.m.). These data were published after the publication of our article [38] and have noticeably lower ambiguity at energies lower 5 MeV (Fig. 2) than do earlier results [44,52–54] and they more accurately determine the general behavior of the *S*-factor at low energies, practically coinciding with early data [51] in the energy range 80–600 keV.

At 1 keV (Fig. 2) the calculated value of the *S*-factor is 0.95 eV b, and calculation results at energies less than 50 keV are slightly lower than data of [5]. In this work, using parameterization of the form

$$S(E_{c.m.}) = S_0 + E_{c.m.}S_1 + E^2_{c.m.}S_2, \qquad (12)$$

a value 2.0(2) keV mb was obtained for $S_0$; for the $S_1$ parameter the value was $1.6(4) \cdot 10^{-2}$ mb and for the S2 $1.1(3) \cdot 10^{-4}$ mb keV-1 was given. The results of this approximation are shown in Fig. 2 by the red solid line and are in a good agreement with the experimental data [5]. In work [5] the results for *S*-factor of the *M*1 transition were also obtained, which lead to its value of 0.008(3) keV mb, that is in 250 times lower than the value $S_0$ = 2.0(2) keV mb obtained in [5]. In the model using by us the cross section of the *M*1 transition equals zero at all.

In [4] the equivalent values were $S_0 = 1.8(1.5)$ keV mb, $S_1 = 2.0(3.4) \cdot 10^{-2}$ mb and $S_2 = 1.1(1.4) \cdot 10^{-4}$ mb·keV$^{-1}$. The results of extrapolation these are given in Fig. 2 by the blue solid line. However, the linear extrapolation of the experimental data in [4,51] down to 1 keV leads to a value of *S*-factor about 0.6(4) eV b, i.e., three times less than in [5]. In addition, the results in [5] have relatively large errors. In order to remove the current ambiguity in data for the *S*-factor of the proton capture on $^3$H, we need new measurements in the energy range from 5–10 up to 30–50 keV.

It is seen from Fig. 2 that at the lowest energies (in the region 1–3 keV) the calculated *S*-factor is practically independent of energy. This suggests that its value at zero energy will not differ from the value at 1 keV. Therefore the difference of the *S*-factor at 0 and 1 keV should not be more than 0.05 eV b and this can be considered as the error of determination of the calculated *S*-factor at zero energy, i.e., $S(0) = 0.95(5)$ eV b.

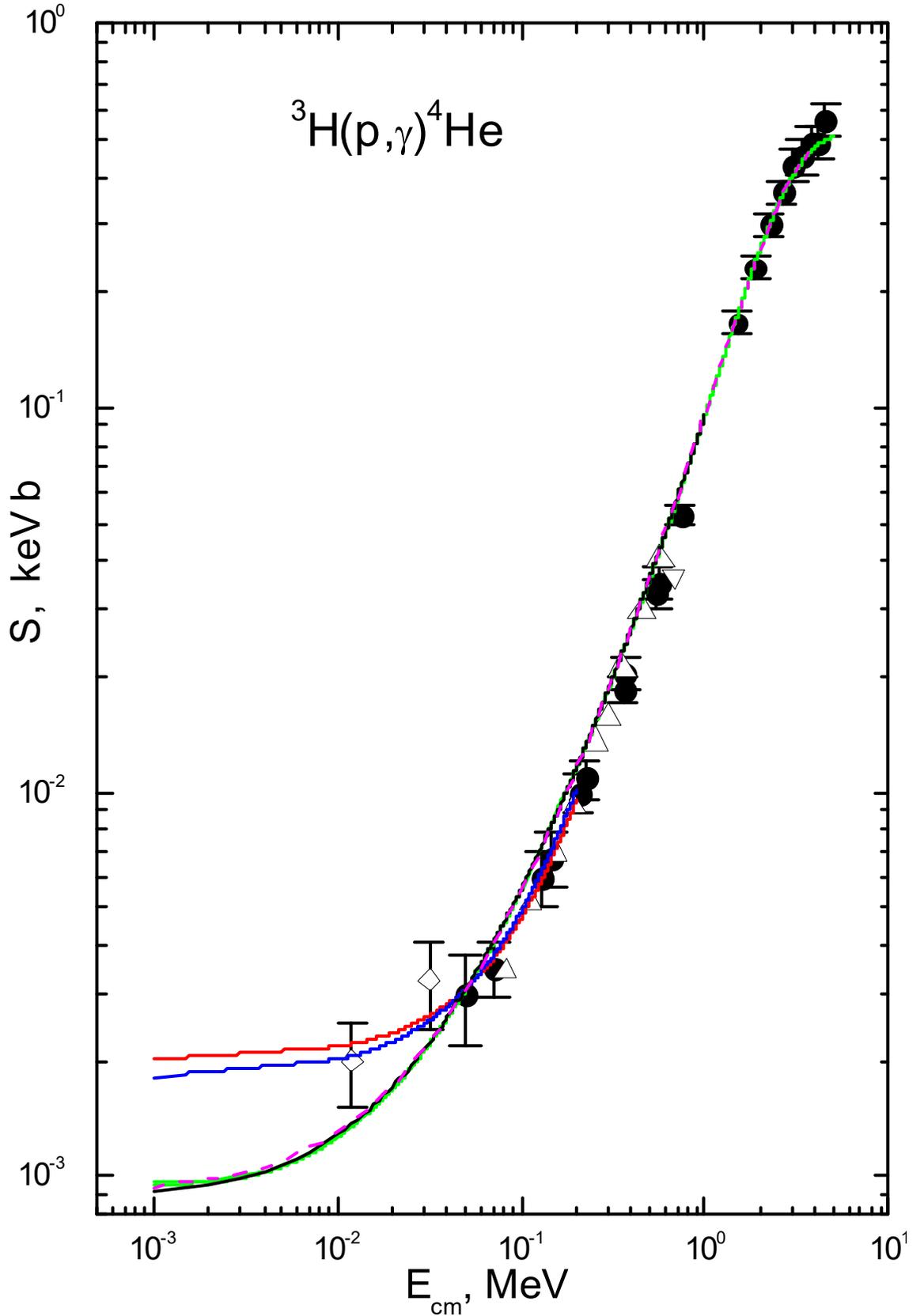

FIG. 2. Astrophysical *S*-factor of proton radiative capture on $^3$H in the range 1 keV–5 MeV. The green line shows the calculation with the GS $^1S$ potential given in (6); the red line shows the results of approximation from [5]; the blue line shows results of the approximation from [4]; the black line shows our approximation. Points show the recalculation of the total capture cross sections [4], given in [5], upward open triangles [51], rhombus [5], downward open triangles [44].

For parameterization of the calculated S-factor in the energy range 1–200 keV the quadratic form (12) can be used and we obtain values for the parameters of: $S_0 = 0.87021$ eV b, $S_1 = 4.086 \cdot 10^{-2}$ eV b keV$^{-1}$, $S_2 = 6.4244 \cdot 10^{-5}$ eV b keV$^{-2}$ at the value of $\chi^2 = 0.23$ at 1% errors of S-factor. The results of such extrapolation are shown in Fig. 2 by the black solid line. This parameterization slightly underestimates S-factor only in the range lower 2–3 keV.

It is possible to use parameterization of the form (12) with addition of the cubic term $E^3_{c.m.}S_3$. In this case at the interval up to 4 MeV for parameters we have found: $S_0 = 0.901194$ eV b, $S_1 = 3.9499 \cdot 10^{-2}$ eV b keV$^{-1}$, $S_2 = 6.94038 \cdot 10^{-5}$ eV b keV$^{-2}$, $S_3 = -1.25131 \cdot 10^{-8}$ eV b keV$^{-3}$ at the value of $\chi^2 = 0.78$ at 1% errors of calculated S-factor. The results of such extrapolation are shown in Fig. 2 by the violet dashed line. Such parameters slightly better describe the calculated S-factor at lowest energies.

For determination of the $\chi^2$ value the usual expression [55] was used

$$\chi^2 = \frac{1}{N}\sum_{i=1}^{N}\left[\frac{S_i^a - S_i^c}{\Delta S_i^c}\right]^2 = \frac{1}{N}\sum_{i=1}^{N}\chi_i^2, \quad (13)$$

where $S^c$ is the initial, i.e., calculated and $S^a$ is the approximated S-factor for the energy denoted by i, $\Delta S^c$ is the error of the initial S-factor, which usually takes equal to 1%, and N is the number of points at summation in the expression given above.

In Fig. 3 the reaction rate $N_A\langle\sigma v\rangle$ of the proton capture on $^3$H is shown (solid blue line). This corresponds to the solid green line in Fig. 2 and is presented in the form [42]

$$N_A\langle\sigma v\rangle = 3.7313 \cdot 10^4 \mu^{-1/2} T_9^{-3/2} \int_0^{\infty} \sigma(E) E \exp(-11.605 E / T_9) dE, \quad (14)$$

where $N_A\langle\sigma v\rangle$ is the reaction rate in cm$^3$mole$^{-1}$sec$^{-1}$, E is in MeV, the cross section $\sigma(E)$ is measured in μb, μ is the reduced mass in amu and $T_9$ is the temperature in units of $10^9$ K which matches our calculation range of 0.01 to 5.0 $T_9$. Integration of the cross sections was carried out in the range 1 keV – 5 MeV for 5000 steps with a step value of 1 keV.

In works [26,29,31] the parametrization of this reaction rate is given

$$N_A\langle\sigma v\rangle = 2.20 \cdot 10^4 / T_9^{2/3} \cdot \exp(-3.869 / T_9^{1/3}) \cdot (1.0 + 0.108 \cdot T_9^{1/3} + \\ + 1.68 \cdot T_9^{2/3} + 1.26 \cdot T_9 + 0.551 \cdot T_9^{4/3} + 1.06 \cdot T_9^{5/3}) \quad (15)$$

The calculation result of such reaction rate is shown in Fig. 3 by the dark dotted line, which appreciably differ from our results, based on the correct description of the astrophysical S-factor in the range from 50 keV to 5 MeV [39] from new work [4] – points in Fig. 2. On the analogy of (15) one can parameterize our calculation results by the analogous form (15), and with other coefficients.

$$N_A\langle\sigma v\rangle = 2.2182 \cdot 10^4 / T_9^{2/3} \cdot \exp(-4.3306 / T_9^{1/3}) \cdot (1.0 + 5.7852 \cdot T_9^{1/3} - \\ - 11.374 \cdot T_9^{2/3} + 12.059 \cdot T_9) - 9.4143 \cdot T_9^{4/3} + 47.332 \cdot T_9^{5/3} \quad (16)$$

with $\chi^2 = 1.3$ at the 1% error of the parameterized reaction rate. Results of such parametrization are shown in Fig. 3 by the red dashed line.

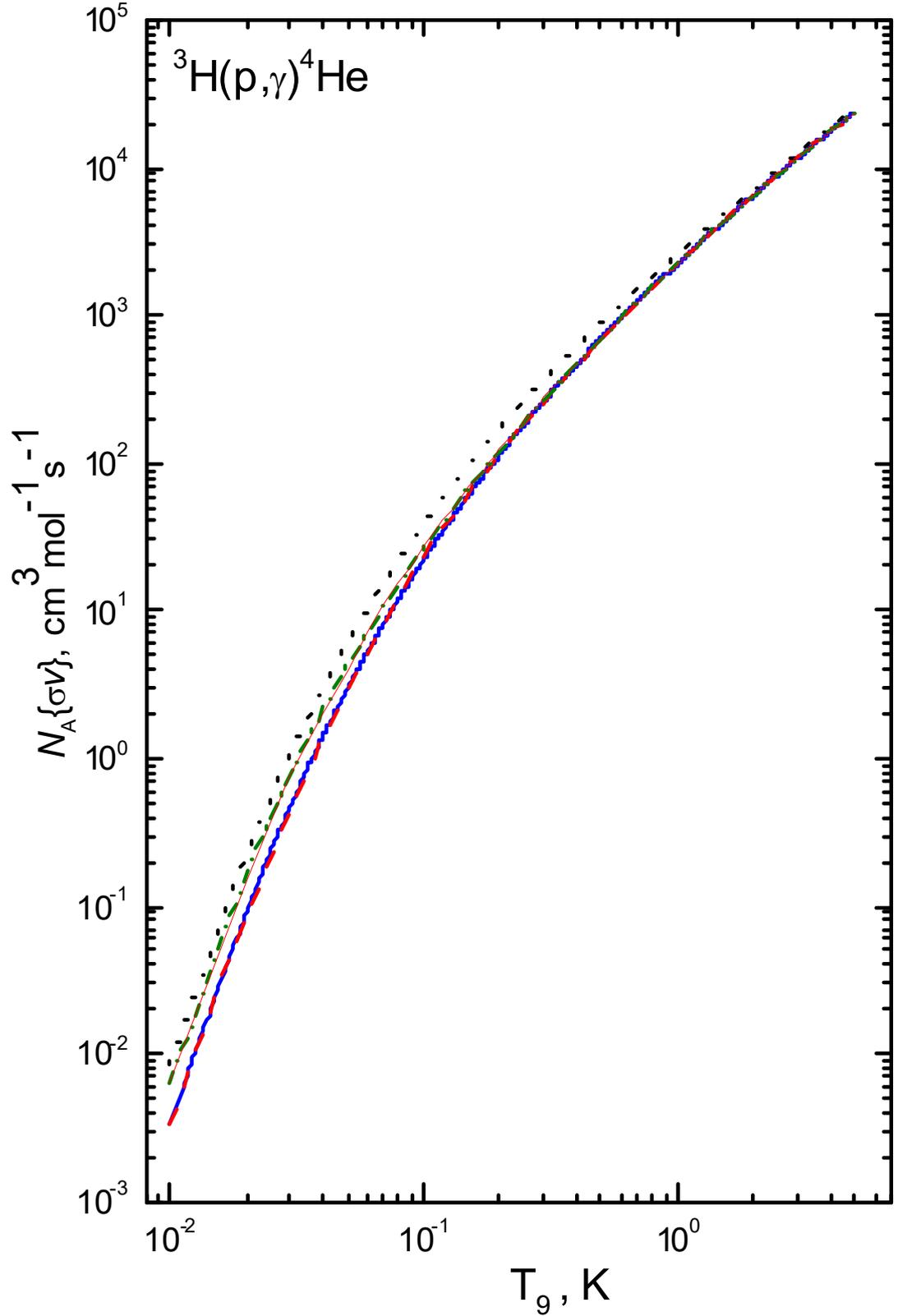

FIG. 3. Reaction rate of the proton radiative capture on $^3$H. Blue line is the calculation results for the GS potential from (6), which correspond to cross sections shown in Fig. 2 by the green solid line.

Possibly, our results for the *S*-factors is slightly underestimated in the energy range lower 50 keV, if proceed from the results [5], shown in Fig. 2 by open rhombus. In order for take into account results of [5] we determine the *S*-factor from parametrization (12) with parameters of work [5] and determine the cross section from the *S*-factor at the range 1 – 50 keV, which at 50 keV coincide with our calculations. Then we change our results by such cross sections in the mentioned energy range and calculate the reaction rate. The results of such calculations are shown in Fig. 3 by the green dotted-dashed line, which from 0.01 up to 0.2 – 0.3 $T_9$ is located slightly higher than our previous results shown by the blue line. However, this result, which completely describes *S*-factor in the range 1 keV – 5 MeV does not coincide with the parametrization [29].

This result of calculations, presented in Fig. 3 by the green dotted-dashed line, can be parameterized by form (15) with parameters

$$N_A \langle \sigma v \rangle = 264.53 \cdot 10^4 / T_9^{2/3} \cdot \exp(-2.5241 / T_9^{1/3}) \cdot (1.0 - 12.5629 \cdot T_9^{1/3} + \\ + 44.0654 \cdot T_9^{2/3} - 35.6503 \cdot T_9 + 59.3843 \cdot T_9^{4/3} + 49.3985 \cdot T_9^{5/3})  \quad (17)$$

with $\chi^2 = 0.1$ at 1% error of the parameterized reaction rate, which shown in Fig. 3 by the thin red solid line.

Since, there was no the most recent measurements for the astrophysical *S*-factor [4,5] at the publishing of the work [26] their parametrization slightly overestimates the reaction rate at low $T_9$, tending to our results at values 3 – 5 $T_9$. Our calculations of the reaction rate given in Fig. 3 by the green dotted-dashed line at the determination of the *S*-factor in the energy range 50 keV – 5 MeV are based on the microscopic MPCM and is not usual parametrization of the experimental data. In the range 1 – 50 keV they empirically take into account the latest data [5] at low energies and can be considered as an improvement of results [26].

## 5. CONCLUSIONS

### *5.1. Nuclear physics*

Thereby, in the framework of considered modified potential cluster model based on the intercluster potentials describing elastic scattering phase shifts and characteristics of the binding state with the potential parameters suggested about 20 years ago [38], on the basis of only the *E*1 transition we succeeded in description of the general behavior of the *S*-factor of the proton capture on $^3$H at energies from 50 to 700 keV. Really, on the basis of analysis of the experimental data above 700 keV [44] about 20 years ago we have done calculations of the *S*-factor for energies down to 10 keV (Dubovichenko 1995). As we can see it now, the results of these calculations reproduce well new data on the *S*-factor, obtained in [4] (points in Fig. 2) at energies in the range 50 keV to 5 MeV.

However the available experimental data on the *S*-factor at 50 keV and below have a low accuracy and significant ambiguity, as it seen from Fig. 2. To avoid these ambiguities, there is a need for new additional and independent measurements of the *S*-factor in the energy range from about 5–10 to 30–50 keV with minimal errors. No new experimental data in this energy range has been forthcoming for more than 10 years [5]. Reliable measurements of *S*-factor at energies of 50 keV–5.0 MeV were made more than 20 years ago [4].

Evidently, modern measurement techniques could reduce error values and obtain more reliable data, especially at the lowest energies. This, in turn, will eliminate the existing ambiguities in determining the reaction rate.

### 5.2. Nuclear astrophysics

The magnitude of the $^3$H(p, γ)$^4$He capture reaction rate calculated in this paper at temperatures from 0.01 $T_9$ up to 5 $T_9$ leads to the conclusion that this reaction will make some contribution to the formation of $^4$He nuclei in the primordial nucleosynthesis of elements in the Universe, especially at higher temperatures around $3.0T_9$–$5.0T_9$. The results obtained for the reaction rate using a simple numerical approximation could be of use in determining the yield of $^4$He in this reaction and thus help estimate the abundance of helium nuclei formed in the primordial nucleosynthesis of the Universe.

We emphasize once again that we could not find other papers with calculations of the astrophysical *S*-factor, in spite of the significance of this reaction represents in terms of some astrophysical problems. Currently available errors of measurements of the astrophysical *S*-factor [5] may significantly affect the value of the reaction rate of the radiative proton capture on $^3$H leading to ambiguities in calculations of $^4$He yield and, ultimately, affect the results obtained for its abundance. Perhaps it is now time to eliminate the existing problems in measuring the astrophysical *S*-factor of reaction under consideration and obtain more accurate results for the reaction rate.

## ACKNOWLEDGMENTS

This work was supported under framework of the financing program "Studying the thermonuclear processes in the Universe" of the Ministry of Education and Science RK through the Fesenkov Astrophysical Institute NCSRT ASC MID RK.

The authors express their deep gratitude to I.I. Strakovsky (GWU, Washington, USA) for discussion of certain issues in the paper.